\documentclass{emulateapj}
\begin{document}
\title{Optical TiO and VO band emission in two embedded protostars: 
IRAS~04369+2539 and IRAS~05451+0037}

\shorttitle{TiO/VO in Emission}

\shortauthors{Hillenbrand et al.}

\author{
Lynne A. Hillenbrand\altaffilmark{1}, 
Gillian R. Knapp\altaffilmark{2},
Deborah L. Padgett\altaffilmark{3},
Luisa M. Rebull\altaffilmark{4},
Peregrine M. McGehee\altaffilmark{3}
}

\altaffiltext{1}{Astrophysics Department, California Institute of
Technology, Pasadena, CA 91125}

\altaffiltext{2}{Department of Astrophysical Sciences, Princeton 
University, Princeton, NJ 08544}

\altaffiltext{3}{Infrared Processing and Analysis Center, California Institute of
Technology, Pasadena, CA 91125}

\altaffiltext{4}{Spitzer Science Center, California Institute of
Technology, Pasadena, CA 91125}

\slugcomment{To appear in the Astronomical Journal}
\accepted{18 November, 2011}


\begin{abstract}
Archival optical spectra from the Sloan Digital Sky Survey of two 
optically faint flat spectrum protostars, IRAS~04369+2539 and IRAS~05451+0037,
show strong emission-line features including -- notably -- clear and broad
emission across several molecular bands of TiO and VO. The molecular emission is
indicative of dense, warm circumstellar gas and has been seen previously 
in only one object: the flat spectrum protostar IRAS~20496+4354 during
a strong optical outburst (PTF 10nvg; Covey et al. 2011). 
The presence of broad molecular emission features in two additional objects 
having similar mid-infrared properties (but not known to be undergoing 
outbursts) could provide new insight into phases of rapid accretion / outflow
at early stages of the protoplanetary disk.  At present, the relevant geometry
and the formation or heating mechanisms responsible for the 
observed TiO / VO cooling emission remain unexplained.

\end{abstract}



\section{Introduction}\label{sec:introduction}

Star formation in nearby molecular clouds -- its efficiency, the 
distribution of initial stellar masses, and the presence of circumstellar
material which is the likely site of planet formation -- has been under
intense observational study for decades (e.g. Evans 1999; Evans et al.
2009).  Young, pre-main-sequence stars (young stellar objects or YSOs)
are discovered and studied at all wavelengths.
Their ultraviolet, optical and near-infrared spectra often show strong 
and variable line emission (Herbig \& Bell 1988; Valenti et al. 2000,2003;
Folha \& Emerson 2001), 
indicating accretion onto the star, with strongly accreting objects 
exhibiting continuum excess over a broad range of wavelengths
(e.g. Gullbring et al. 1998; Fischer et al. 2011, Furlan et al. 2011).
Although both embedded or ``cocoon" stars (e.g. Becklin \& Neugebauer, 1967) 
and optically visible stars with excess mid-infrared emission 
due to circumstellar dust (e.g. Mendoza, 1966; Low et al. 1970;
Strom et al. 1971; Cohen, 1973)
had been discovered using ground-based techniques, 
it was infrared imaging by the {\it InfraRed Astronomy Satellite} ({\it IRAS}; 
Neugebauer et al. 1984) that revealed large numbers of deeply embedded, 
optically obscured YSOs, and also the prevalance of dust excess.  More recent
infrared surveys using the {\it Spitzer} (Werner et al. 2004) and 
{\it Wide-field Infrared Survey Explorer} ({\it WISE}; Wright et al. 2010) 
space observatories, with their much greater sensitivity,
spatial resolution and astrometric accuracy, have detected thousands 
of YSOs in nearby molecular clouds (e.g. Evans et al. 2009; Rebull et al. 2010),
identifying objects with masses down to, and below, that of the lowest-mass
main sequence star and enabling detailed studies of planet-forming disks.

A third defining property of YSOs, in addition to emission-line spectra and
continuum excess from ultraviolet to infrared, is time-variability. 
This phenomenon is manifest at low levels on day-to-week time scales, 
attributed to rotation/star spots (regular variability) as well as 
accretion-related jitter (irregular variability) as discussed by e.g.
Herbst et al. (1994),
Carpenter et al. (2001) and Stassun \& Wood (1999).  More exotic variability
at amplitudes of 3-5 magnitudes occurs on few year to 10,000 year (estimated) 
time scales and is attributed to occasional accretion-related outbursts 
(the EX Lup and FU Ori variables).  Synoptic studies of wide areas of
sky are now in regular operation, and have identified large
numbers of variable young objects of many different types, 
including outbursting protostars. 

Recently, Palomar Transient Factory (PTF; Law et al. 2009)
observations of the North America Nebula have reported the discovery of 
a strong ($> 5^m$) outburst from the source PTF~10nvg, which is 
associated with IRAS~20496+4354;
detailed photometric and spectroscopic follow-up observations are reported in 
Covey et al. (2011) and Hillenbrand et al. (2012, in preparation). 
The object's initial optical and near-infrared spectra were characterized
by strong emission lines in hydrogen, helium and various metallic elements,
which is not unusual for a young star.  However, the object also exhibited
the heretofore unique feature of broad, strong emission in optical spectra
that was associated with bands of TiO and VO. 

As noted in Covey et al., such optical molecular emission had been only hinted 
at in the previous astronomical literature, with
TiO emission in one or few {\it bandheads} reported in 
several luminous B[e] stars and several evolved and dusty objects. 
For young stars, a similar phenomenon had been mentioned in just the 
one case of V1057 Cyg (Herbig, 2009; the weak emission 
in the 7054, 7087, and 7125~\AA{} heads that Herbig mentions
is confirmed in our own spectra of this object).  The
difference between these previously identified objects 
and the cases of TiO/VO emission in IRAS~20496+4354 and the two young stars
we present here, is in the morphology of the emission.  
These three systems differ from those previously noted in the literature
in both the breadth of the emission spectrum 
(broad, full molecular band vs a few band-heads only) 
and the emission intensity (far stronger).  
TiO and VO are usually indicative of warm and dense molecular material with
lines from these molecules seen prominently in absorption in all
late type (M to early L) stars.  

The present paper describes archival observations from the {\it Sloan
Digital Sky Survey} (SDSS; York et al. 2000) of two more protostars, 
in addition to the case of IRAS~20496+4354, that
show broad TiO and VO emission bands in their optical spectra: 
IRAS~04369+2539 and IRAS~05451+0037. 
The available data suggest, however, that unlike IRAS~20496+4354
these objects do not exhibit evidence of large amplitude
photometric variability that might be associated with a recent outburst event.
Section 2 presents the {\it SDSS} photometry and spectroscopy. 
A description of the two YSOs 
and their broad-band spectral energy distributions is given in Section 3.
The results and explanatory hypotheses
are discussed in Section 4 and the conclusions presented in Section 5.


\section{Observations: SDSS Photometry and Spectroscopy}\label{sec:data}

The optical photometry and spectroscopy were obtained as part of SDSS
imaging of about a quarter of the
sky in five bands covering the optical wavelength range: $u$ (centered at
about 3543 \AA{}), $g$ (4770 \AA{}), $r$ (6231 \AA{}), $i$ (7625 \AA{}),
and $z$ (9134 \AA{}). Magnitudes are measured on the AB system (Oke \&
Gunn 1983) to an accuracy of about 0.01 mag for point sources brighter
than about 18.5 mag (Ivezi\'c et al. 2004) and to a 5$\sigma$ limiting 
magnitude of about 21.5.
The spectra are measured by a pair of fiber-fed
multi-object spectrographs (Uomoto et al. 1999) with a resolution of 1800
-- 2100 over a wavelength range of 3800 -- 9200 \AA{}. 
The spectral data reduction includes optimal extraction of the data, 
flat--fielding, wavelength calibration using arc and night--sky spectra,
sky subtraction using the spectra observed through fibers assigned to
blank sky locations, flux calibration relative to the spectra of subdwarf
F stars selected by color, and correction of the wavelength scale
(measured in vacuum wavelengths) to the heliocentric standard of rest. 
Table 1 summarizes the {\it SDSS} photometric and spectroscopic
observations of the two objects of interest. 

The optical spectra for IRAS~04369+2539 and IRAS~05451+0037 were obtained during
small surveys with the {\it SDSS} spectroscopic instrumentation of objects 
selected from {\it SDSS} photometric observations of parts 
of the Taurus and Orion star-forming
regions. These data, not part of the main {\it SDSS} survey, 
were obtained for test purposes (Taurus) and to 
search for red YSOs (Orion - see McGehee et al. 2004). 
Indeed, the {\it SDSS} equatorial scans through part of the Orion
constellation, which cover much of the region around the
reflection/emission nebulae NGC 2068 and NGC 2071,
were among the first data taken by the {\it SDSS} camera and were released 
publicly in association with their description in Finkbeiner et al. (2004). 

For IRAS~04369+2539, one {\it SDSS} photometric and one spectroscopic
measurement are available. The star is not detected in the $u$ and $g$ bands.
The spectrum (Figure 1) 
provides a second set of $r$ and $i$ magnitudes, which were
synthesized by convolving it with the {\it SDSS} filters;
again, there is no detectable flux in the wavelength range of the $g$ band.
IRAS~05451+0037 was imaged six times, 
in three groups of observations separated by about six years, with 
separations of 1- 2 days within these two groups. 
{\it SDSS} also obtained one spectrum (Figure 1) of IRAS~05451+0037 from which
$g$, $r$ and $i$ magnitudes were synthesized. 
For both objects, photographic photometry from
the Palomar Observatory Sky Surveys as listed in the USNO-B
catalog (Monet et al. 2003) is also included in Table 1.
These optical magnitudes show very little variability for either object.

Figure 1 shows the {\it SDSS} spectra for the two YSOs.  In Figure 2
the spectra are dereddened as per the discussion below, and 
compared with the spectrum of an undistinguished dM2 emission-line star in the
Orion region as well as with the spectrum of IRAS~20496+4354 (PTF~10nvg)
from Covey et al. 2011. 
The Taurus star-forming region has very few OB stars and 
almost no diffuse HII region emission; 
the Balmer, and any nebular (e.g. [\ion{N}{2}] and [\ion{S}{2}]),
emission lines in IRAS~04369+2539 would thus be reliably measured. 
IRAS~05451+0037, on the other hand, lies near NGC 2071 with hot stars 
and extended diffuse HII region emission; 
Balmer emission here would be suspect, therefore. 


\section{The Molecular Emission Objects}\label{sec:objects}

\subsection{Optical Spectra}

The unusual nature of the optical spectra of IRAS~04369+2539 
and IRAS~05451+0037 was noted during current studies by our group
of the star-formation process in the Taurus and Orion regions 
based on data from the {\it Spitzer} and {\it WISE} mid-infrared maps 
of these regions, and their intersection with available {\it SDSS} data.  
While the mid-infrared photometry can be used to select objects 
bearing significant mid-infrared excess emission, indicative of dusty disks,
the optical data can be examined for emission-line stars and those with 
significant blue excesses, indicative of accretion processes.  

Specifically, two emission-line
objects were found in the {\it SDSS} spectroscopic database with spectra
similar to that of IRAS~20496+4354 (PTF~10nvg): 
IRAS~04369+2539 (SDSS~J043955.76+254501.9) and 
IRAS~05451+0037 (SDSS~J054744.99+003841.2).
The two sources exhibit (Figures 1 and 2) not only strong atomic emission
lines, in particular strong 
\ion{Ca}{2} triplet lines at about 8498, 8542, and 8662\AA{} 
and weaker \ion{O}{1} at 8446 ~\AA~ along with H$\alpha$, 
indicating active accretion (Azevedo et al. 2006;  Kwan \& Fischer 2011), 
but both spectra also show the presence of optical bands of TiO and VO
{\it in emission}.  IRAS~05451+0037 shows a more
complex spectrum, with [\ion{O}{1}] at 6300,6363\AA{}, [\ion{S}{2}] at 6717,6731\AA{}, 
and perhaps several weak [\ion{Fe}{2}] lines also in emission. 

Further, as shown in Figure 3 which displays close-ups
of the spectra from Figure 1 in the wavelength regions containing the 
\ion{Na}{1} D, \ion{K}{1} and \ion{O}{1} lines, 
these metallic lines are seen {\it in absorption against the molecular
band emission}.  Figure 3 also shows the region of the H$\alpha$ line, 
which is strong in both objects though, as noted above, the line shape
and strength is unreliable for 
SDSS~J054744.99+003841.2 (IRAS~05451+0037). 
Again for comparision, the optical spectrum of IRAS~20496+4354 (PTF~10nvg)
from Covey et al. (2011) is included in Figure 3.

\subsection{Context from Previous Literature}

SDSS~J043955.76+254501.9 is coincident in position with IRAS~04369+2539,
the infrared source adjacent to the large reflection nebula IC~2087;
it is also known as IC 2087-IR and Elias 3-18, located in the Heiles Cloud 2
region of Taurus.  The star was classified by Elias (1978) 
as an early type object based on luminosity arguments,  roughly a B5 star 
if on the main sequence or a later type if still pre-main sequence.
White \& Hillenbrand (2004) attempted 
spectroscopic classification from analysis of a high dispersion 
optical spectrum, settling on a ``K:" spectral type based on the  
\ion{K}{1} absorption 
and possible hints of \ion{Ti}{1} absorption near 8425 and 8435 \AA, 
both consistent with a mid-K star. Referring to the second page of their 
Figure 2, the \ion{K}{1} absorption now can be recognized as having 
a narrow component at zero velocity plus a broad blueshifted component, 
as seen in the SDSS spectrum of our Figure 3, while
emission from individual TiO bandheads can indeed be seen in their
right panel at 8432, 8441, and 8451 \AA. 
The source also shows broad \ion{Ca}{2} and H$\alpha$ emission, the latter with a 
sub-continuum P-Cygni absorption superposed on the emission
but no forbidden line emission (their Figure 7 and Table 4).
More recent spectroscopic study of the source by Connelley \& Greene (2010) 
at near-infrared wavelengths revealed no photospheric absorption lines
and therefore no spectral type assignment was made.  
The source shows strong emission
in the CO vibrational band at 2.3$\rm \mu m$, as do about
25\% of the non-FU Ori YSOs in the Connelley \& Greene survey of Class I stars.
Paschen and Brackett emission is also prominent, and the source exhibits
a \ion{He}{1} $\lambda$1.083$\rm \mu m$ profile that is reminiscent of the H$\alpha$ 
profile described above.  
As is the case for IRAS~20496+4354 (PTF~10nvg; Covey et al. 2011), while  
the CO overtone at 2.3$\rm \mu m$ is in emission in IRAS~04369+2539,
the CO fundamental at 4.6$\rm \mu m$, which measures cooler gas,
is in absorption (Shuping et al. 2001).
There is no forbidden-line or $H_2$ emission
in the Connelley \& Greene (2010) spectrum,
consistent with the lack of outflow signatures in optical spectra.
Although optical and infrared spectra lack the usual indicators of jets and
outflows, the redshifted lobe of a 
molecular outflow was detected by Heyer et al. (1987) 
and nearby Herbig-Haro objects 395A and 395B were observed by 
Gomez et al. (1997; notably outside the lobe of 
the redshifted molecular gas); this may 
suggest that the outflow was stronger in the past.
Both White \& Hillenbrand (2004) and Connelley \& Greene (2010)
derived an extinction (A$_V$) around 18 mag from de-reddening 
of near-infrared colors;
this is almost identical to the extinction value first derived by Elias (1978).
The gas-phase and solid-phase circumstellar disk/envelope chemistry 
is well-studied for this source (e.g. Tielens et al. 1991; Chiar et al. 1995;
Shuping et al. 2001; Palumbo et al. 2008; Furlan et al. 2008;
Whittet et al. 2010).

SDSS~J054744.99+003841.2 is an extremely red star to the 
northwest of NGC 2071, in the NGC 2071-North region, and is 
very likely the same object as IRAS~05451+0037 despite being
outside the formal error ellipse.  The {\it IRAS} source has been observed 
in several molecular line surveys but is little studied otherwise and, 
specifically, there are no data previously reported at
optical wavelengths; the spectrum obtained by {\it SDSS} is, as far
as we are able to determine, the first published for this object. The
region contains the emission-line star LkH$\alpha$319 several arcmin
to the east, and approximately the same distance to the west are
LkH$\alpha$316 / HBC 510 and LkH$\alpha$316c / HBC 511 along with
Herbig-Haro objects 473 and 474 on either side of 
SDSS~J054744.99+003841.2 (Aspin \& Reipurth 2000), and a large
bipolar molecular-line outflow (Iwata et al. 1988; Goldsmith et al. 1992).  
As is the case in the IRAS~04369+2539 region, the molecular outflow and the
HH objects are rather mis-aligned, perhaps
indicating a stronger or differently directed flow in the past. 
2MASS images show nebulosity closely associated with the source
at K and H bands, which is not seen at shorter wavelengths. 
WISE images indicate that the optical/infrared source we have identified here
is the brightest in the region at mid-infrared wavelengths, though by 25 $\mu$m
the flux from LkH$\alpha$316 and LkH$\alpha$316-neb (designations from 
Reipurth \& Aspin) is also quite strong. We believe that the presence 
of several sources in close proximity having quite different 
spectral energy distributions explains the positional offset of
IRAS~05451+0037 from the brightest peak that is reveled 
in higher spatial resolution mid-infrared data as 
coincident with SDSS~J054744.99+003841.2.

\subsection{Spectral Energy Distributions}

The broad-band spectral energy distributions (SEDs) 
for IRAS~04369+2539 and IRAS~05451+0037 were constructed from catalog
data obtained with {\it SDSS, 2MASS, WISE, IRAS, Akari, Planck}
and literature data from {\it Spitzer} 
and the SCUBA camera on the James Clerk Maxwell Telescope, as detailed below. 

The {\it Spitzer} IRAC data for IRAS~04369+2539 are saturated, 
as reported in the catalog of Rebull et al. (2010), but
Luhman et al. (2010) report flux densities at 5.8 and 8.0 $\rm \mu m$
that are brighter than the Rebull et al. lower limits. 
Rebull et al. (2010) provide a {\it Spitzer} MIPS 70$\rm \mu m$ flux density 
and a 160 $\rm \mu m$ upper limit.  Flux densities at 450 and 850$\rm \mu m$ 
are reported by Andrews \& Williams (2005) and in
the catalog by DiFrancesco et al. (2008). 
IRAS~05451+0037 does not appear in the {\it Spitzer} catalogs of YSOs in this
region discussed by Flaherty \& Muzerolle (2008) and Fang et al. (2009) and it
appears that the data for this bright object are saturated in
all {\it Spitzer} observations. 
Both {\it IRAS} objects are detected by the Infrared Camera
as well as the Far Infrared Surveyor on {\it Akari} 
(Kawada et al. 2007; Murakami et al. 2007; Onaka et al. 2007).
The {\it Planck} Early Release Compact Source Catalog (Planck Collaboration 2011a,b)
lists a submillimeter source near the position of IRAS~05451+0037. 

The SEDs (Figure 4) of the two TiO / VO emission sources
have rather similar shapes, with most of the energy emitted 
between wavelengths of about 3 and 100 $\mu$m.  
While the SEDs are flat or rising in flux density units, 
they are not steep enough for Class I designation.
IRAS~04369+2539 is a Class II type YSO in the usual 2 to 25 $\mu$m
definition ($\alpha = -0.42$
where $\alpha > -0.3$ is considered flat spectrum and 
$\alpha > +0.3$ is considered Class I; Greene et al. 1994) while
IRAS~05451+0037 is also a flat spectrum type YSO ($\alpha = +0.16$).
As discussed below where we model the SEDs in detail, and 
consistent with the usual interpretation of Class I and flat-spectrum sources,
for IRAS~05451+0037 there is likely a significant
circumstellar envelope from which material is still accreting 
onto an extended, massive disk. IRAS~04369+2539 on the other hand,
consistent with the usual interpretation of Class II sources,
is likely more disk dominated rather than envelope dominated. 
For context, the spectral slope of IRAS~20496+4354 -- the only other
broad TiO / VO emission source known -- is $\alpha = +0.22$, steeper
than both IRAS sources discussed here.

The SEDs of Figure 4 both show peaks
at 4.3 $\rm \mu m$ (in WISE band 2) but this is not likely to be real because
of non-linearity effects in this band in the currently public
versions of the reduced data, affecting bright sources. 
It is also possible that the emission at long wavelengths ($>$100 $\mu$m) 
contains contributions from the extended dark clouds in which both
of these sources are embedded, due to the large beam sizes of the
telescopes.  Ignoring both the long-wavelength flux and the 4 $\mu$m
flux, and using the assumption of isotropic emission, the integrated spectra
give bolometric luminosities of about 7 $\rm L_{\odot}$ for IRAS~04369+2539
(assumed distance of 140 pc to the Taurus star-forming region;
 Loinard et al. 2004, Torres et al. 2009)
and 90 $\rm L_{\odot}$ for IRAS~05451+0037 (assumed distance of 400 pc
to NGC 2071; Gibbs 2008). 

The SEDs were modeled using 
the code and results described by Robitaille
et al. (2006, 2007), Robitaille (2008), and Wood (2008). 
In the case of both stars, it
is impossible to get an excellent model fit to all of the data, so as above,
the data for wavelengths beyond 70 $\rm \mu m$(IRAS~04369+2539) and beyond 
100 $\rm \mu m$ (IRAS~05451+0037), as well as the 4$\rm \mu m$ WISE
measurements, were not used in the fits. The optimum models for both objects
are indeed stars surrounded by relatively massive disks, with 
the optical flux dominated by scattered light and requiring 
significant foreground extinction. 
The characteristics of representative (but non-unique)
models are listed in Table 2.  We quote the ``best-fit" model parameters
but emphasize that in the face of high $\chi^2$ values ($>$45)  
and given the sparse sampling of the SED combined with the many 
geometric, dust, and radiative parameters that would be important to vary, 
the fit parameters can not be considered as well determined.  It is for this
reason that we do not show the SED fits in Figure 4.
However, among the ``top ten" fits for each source,
the representative model envelope mass in IRAS~05451+0037 is much 
higher than that for IRAS~04369+2539, 
consistent with the differences in spectral slope.
The fitted models do not require any envelope accretion.
The disk masses are the same to within a factor of two.
The disk accretion rates are consistent with those inferred for Class I 
sources and higher than those typically measured for Class II sources. 
Significant inclination (i $>70-80^\circ$) is a consistent feature of the models.

The model fitting includes a solution 
for the foreground extinction.  These derived extinction values 
were used to correct the optical spectra shown in Figure 2.
The inferred extinction from the modelling procedure
towards IRAS~04369+2539 is lower 
than that inferred from other methods (A$_V$=5-9 vs the 18 mag mentioned 
earlier).  The lower values are driven by a model including dust geometry 
and physics, grain absorption, re-emision and scattering, 
and system orientation, that is then reddened by a foreground 
screen to fit the SED.  The quoted A$_V$ is the foreground extinction,
which is somewhat biased by matching the short wavelength points.  
The higher value comes from consideration of near-infrared photometric
or spectrophotometric data only, relative to a presumed underlying star,
and may not correctly separate the contributions of foreground vs
circumstellar extinction nor account for circumstellar emission. 
The model foreground extinction towards IRAS~05451+0037 (A$_V$=6-10 mag) 
is comparable  to that for IRAS~04369+2539 and, as noted earlier, 
no previous estimates exist in the literature for this source.


The $Q_{riz}$ reddening-invariant index, defined as
$Q_{riz} = (r-i) - \frac{E(r-i)}{E(i-z)}(i-z)$ as in McGehee et al., (2004), can in principle be used to infer both intrinsic optical colors and foreground extinction. Under the assumption of a Fitzpatrick (1999) reddening law, the multiplier to $(i-z)$ has the value of 1.349 or 1.091 for $R_V$ = 3.1 or 5.1,
representing a standard dust model or that appropriate for the 
larger grains found in dense molecular clouds (Schlafly \& Finkbeiner, 2010). 
Comparison of the computed $Q_{riz}$ values (Table 1) against those for representative spectral types along the stellar locus in SDSS colors
(Covey et al., 2007) fails to produce an extinction value and, instead,
suggests that both IRAS sources are veiled in the optical. 
The extent of any inferred veiling component is dependent upon 
the assumed $R_V$ value.  For a dust model with $R_V = 3.1$, 
the $Q_{riz}$ index is $>$ 0 for all spectral types, and has a value 
of 0.16 for M0 V stars; both sources appear to have an excess 
optical continuum, especially at the (shortest) $r$ band. 
In the case of $R_V = 5.1$ the reference value at M0 V is $Q_{riz}$ = 0.25 
and thus the evidence for excess optical continuum is weaker.
The optical veiling may arise from a combination of phenomena 
including a magnetospheric accretion shock, hot gas in the circumstellar disk 
environment, and scattered light.  As noted above, the SED modelling procedure 
produced evidence for significant amounts of scattered light at these
wavelengths, and if the disk is accreting at the rates indicated by these
same models, the presence of accretion-induced veiling is not surprising.

\section{Discussion}

\subsection{TiO / VO Emission in Young Stellar Objects}

The most striking feature of the optical spectra of 
IRAS~04369+2539 and IRAS~05451+0037 
is the broad emission seen throughout several TiO and VO band regions.  
The basic physics of molecular radiation 
involves some combination of rotational and vibrational (bending/stretching) 
plus electronic transitions, with numerous possible modes but 
all following selection rules.
Excitation of molecular emission is usually taken as being collisionally driven,
though photo-excitation  methods can also be relevant (see below). 
Molecules are present when there is gas of the appropriate 
temperature and density. 
The specific presence of TiO and VO is indicative 
of warm ($\sim$1500-4000 K; Lodders 2002; Ferguson et al. 2005; Sharp \& 
Burrows 2007) gas at high density ($> 10^{10} cm^{-3}$).  The resolution 
of the existing optical spectra is not high enough to attempt to constrain 
the temperature using a proper vibrational/rotational model 
for the molecules, however. Similarly, we are unable to assess the 
individual line widths so as to ascertain the kinematics of the emission region.  

Questions that arise concern how the TiO and VO molecules form in a
circumstellar environment,
how they are heated so that cooling produces the observed
molecular emission, and finally why the phenomenon is so rare.  
With emphasis on the most abundant HCNO composition molecules and ions,
Semenov (2011) provides a high level overview of our knowledge 
of molecular species in circumstellar disks and Visser et al. (2011) present 
the detailed chemical history from the collapse phase to the disk. 
While Ti and V chemistry is not generally considered because of the low
elemental abundances, constraints on the TiO/VO emitting region come from
the consideration that these molecules must be among the first condensates in 
forming circumstellar disks to freeze out, given the relatively high
condensation temperatures. Thus, their observed presence in the gas phase
is quite unexpected unless they are able to convert from solid-phase
back to gas-phase, or if they can form anew from their atomic components 
within the circumstellar environment.  


%

Turning now from formation to heating,
in the canonical geometry of a star/ disk/ envelope system there is not 
the needed level of temperature inversion 
in stellar photosphere heated circumstellar disk models 
from the disk mid-plane to the disk surface 
that would be necessary to produce the
TiO / VO emission.   However, if in addition to the expected
spectral energy distribution from the stellar photosphere there is 
sufficiently high additional UV irradiation of the disk surface, 
or x-ray photons that penetrate into the disk, then
molecular emission could in principle result from a super-heated
disk atmosphere.  These heating mechanisms are indeed implicated 
in the limited results to date on warm molecular gas in young disks\footnote{
We note that the majority of molecular gas studies are of
colder gas components in the outer disk at tens to hundreds of AU
having temperatures of a few tens of K (mostly sub-mm and far-ir studies)
to a few hundred K (mostly mid-infrared to near-infrared). Hotter gas
from the chromosphere, coronal, and accretion shock regions with
10$^4$ K to 10$^6$ K temperatures is probed by UV to x-ray studies.}, 
notably the photo-excitation of H$_2$ mainly by FUV line emission
(Herczeg, 2004; France et al. 2007, 2010, 2011). 

For the H$_2$ molecule specifically,
temperatures in the $\sim$1500-3500 K range have been derived, 
similar to those needed to produce the TiO/VO emission seen
in our sources.  The warm H$_2$ gas is postulated to arise 
in the disk atmosphere, within one or a few AU of the star.  
Most other detected molecules 
(Salyk et al. 2011ab; Carr \& Najita 2011; Doppmann et al. 2011)
are of HCNO composition and have cooler temperatures, only 500-1500 K, though 
some warmer OH has been detected (Najita et al. 2010; Carr \& Najita 2011) 
with $\sim$4000 K.
This molecular emission may also arise in the inner disk region with, again,
UV heating implicated (here, of H$_2$O which dissociates to OH). 
Because of their self-embedded nature, any such high energy emission
would be undetectable in the sources discussed here, so we can not test
directly this explanation for the instigation of warm TiO/VO emission
in IRAS~04369+2539,  IRAS~05451+0037, or IRAS~20496+4354 (PTF~10nvg).

However, a sudden change in temperature or
perhaps in density could result in not only TiO and VO formation
but also heating.  For PTF~10nvg,
the optical/infrared outburst is the likely culprit for 
increased heating of the disk, perhaps pushing the ionization / dissociation
zone either radially outward or vertically further into
the disk at a given radius, from the surface heating. 
The TiO and VO could then form either via
evaporation / desorption from grains, or anew as the gas cools.
The TiO/VO formation mechanism is not as easily explained in 
IRAS~04369+2539 or IRAS~05451+0037 since these sources do not appear 
to have changed their recent brightness significantly (see section 4.3 below). 
However, extreme UV or X-ray flaring that is not disk accretion-related
could be another option.

Another possibility is to form the broad TiO/VO emission ``continuum'' in the same
place as the \ion{K}{1} absorption seen against it: within the outflow.
As the outflow mechanically clears the near-circumstellar environment,
new material is exposed to the
stellar-plus-accretion radiation field where, again, TiO and VO could
form as the gas cools.
The warm emission could arise in a dense part of the outflow, perhaps near 
its base, as material is lifted out of the disk to heights where it can be 
illuminated directly by UV and x-ray radiation,
coming from either the star (chromosphere / corona region) itself,
or the accretion shock.  
In all three of the {\it IRAS} sources discussed here,
low excitation \ion{Na}{1} D and \ion{K}{1} $\lambda$7665,7699, 
and the higher excitation \ion{O}{1} $\lambda$7773 triplet are seen in
absorption against the broad TiO/VO band emission (Figure 3).
We conclude that the TiO/VO emitting region in these objects must, 
therefore, be within or below the outflow zone.
In the case in IRAS~20496+4354 (PTF~10nvg), higher dispersion data
(Covey et al. 2011) show that these lines were clearly blueshifted,
consistent with their formation in a wind. There is some hint even in the lower
spectral resolution data of Figure 3 for such blueshifts in all three sources.  

\subsection{Connection to Other Warm Molecular Emission in Young Stars}

Glassgold et al. (2004) and Najita et al. (2011) have explored theoretically
the origin of various species of molecular emission in the upper atmospheres 
of disks, where conditions would be similar to the temperature inversion 
scenario suggested above as a possibility for the observed TiO/VO emission.  
These detailed studies of disk heating/cooling physics and chemistry do not 
include TiO and VO cooling, however.  Muzerolle et al. (2004) describe
LTE consideration of molecules including TiO in their disk models, 
but do not discuss any molecular emission (only absorption
at high accretion rates).


The morphology of the molecular emission that we see at optical wavelengths 
bears a resemblance to the broad H$_2$O emission reported by
Carr et al. (2004) at infrared wavelengths,
in a young outflow-driving source (SVS~13).  The temperatures 
inferred were warm ($\sim$ 1500 K) but somewhat lower 
than those seemingly required for the optical TiO / VO emission. 
Such temperature differences would correspond in a simple disk model
to differences in the radii at which the emission arises.
Notably, however, SVS~13 also exhibits warm CO emission, which is 
seen more frequently in young stars than either broad H$_2$O or TiO / VO.

The CO overtone region near $2.3 \mu$m has temperature and density conditions 
(T $>$1000-2000 K; n$> 10^{9} cm^{-3}$) similar to those for TiO and VO. 
Molecular species, including CO, TiO, VO, and H$_2$O, are all prominent 
in the absorption spectra of M-type photospheres.  Near-infrared 
CO emission is not uncommon among high accretion rate young star/disk systems,  
and high resolution observations of the CO bandhead have been modeled as 
originating in the inner few AU of a rotating disk (Najita 1996), in 
the magnetospheric infall region from the disk to the star (Mart\'{\i}n 1997), 
or in the base of the outflow (Carr 1989). In all of these models
the inner disk region is likely the physical region of the emission, though
the suggested kinematics are different.
TiO and VO molecules may probe regions of similar or perhaps 
slightly hotter and denser gas relative to CO.

In at least two prominent CO overtone 
emitters, DG Tau and EX Lup, the CO bandhead is time variable with 
absorption displayed during quiescence and emission exhibited during 
periods of enhanced accretion (Biscaya et al. 1997; Lorenzetti
et al. 2009; K\'osp\'al et al. 2011).
The TiO/VO emission in IRAS~20496+4354
(PTF~10nvg; Covey et al. 2011) was clearly variable on several month
time scales (as noted earlier this source also showed CO overtone emission 
during this time period, though CO fundamental {\it absorption}).  Repeated
observations of IRAS~04369+2539 and IRAS~05451+0037,
in which the TiO/VO emission
is announced for the first time here, are encouraged.

\subsection{Connection to Variability}

The assembled optical magnitudes in Table 1 are sparsely sampled.
While the SDSS photometry and spectrum-synthesized magnitudes
were obtained over only about
ten years, the USNO-B (Monet et al. 2003) magnitudes date from the
1950's, and together indicate little long term variation in the
two {\it IRAS} sources discussed here when the errors of the 
photographic photometry are taken into account.  If anything, the long term
trend for both sources may be one of fading rather than the
brightening observed to reveal TiO / VO emission in IRAS~20496+4354.
The shorter time scale variations in the optical photometry from epoch to epoch
are characteristic of the levels seen in YSOs due to a combination of
starspots or small variations in accretion or variable extinction.  
The apparent variations in $Q_{riz}$ in Table 1 
are at a similar significance to those in the photometry. For both,
the variations are at the 3-10 $\sigma$ level.
Further multi-wavelength synoptic observations
would perhaps better illuminate the variability trends in these objects.

%
%

Further, the broad-band spectral energy distributions 
in Figure 4 are constructed from data taken over more than 25 years, 
and suggest no dramatic photometric variation during that time. 
This situation can be contrasted with the large amplitude variation seen in the 
broad-band SED of IRAS~20496+4354 / PTF~10nvg (Covey et al., 2011)
during which time the TiO/VO broad emission was observed.
These considerations suggest that the rapid accretion
onto IRAS~04369+2539 and IRAS~05451+0037 
as indicated by strong optical atomic emission features 
and probably related to the broad optical TiO/VO emission 
(as well as the infared CO emission; see Appendix)
has been underway for at least decades.
The lack of strong variability is particularly notable given the 
generally large optical and infrared photometric amplitudes 
exhibited by known CO overtone emitters.

\section{Conclusions}

(1) Archival SDSS optical-wavelength spectra for two embedded protostars,
IRAS~04369+2539 (IC2087-IR in the Taurus star forming complex) 
and IRAS~05451+0037 (in the NGC 2071 N 
star formation region) show strong emission-line spectra indicative of
accretion, likely blueshifted absorption lines related to a wind, 
and -- of particular interest -- 
broad emission throughout the optical TiO and VO
bands. Only one other object, IRAS~20496+4354/PTF~10nvg, is previously
known to exhibit similarly strong, broad optical molecular emission.

(2) The SEDs of all three sources are characteristic of 
flat-spectrum type protostars, for which extended massive disks 
and some envelope components are generally
required in order to model their photometry.

(3) Unlike the case of IRAS~20496+4354, the TiO/VO emission does not
appear to be associated with a strong recent outburst
in IRAS~04369+2539 or IRAS~05451+0037. 

(4) The presence of TiO/VO requires high densities and moderate temperatures.
The emission results possibly from enhanced heating (e.g. UV or x-ray) 
of pre-existing gas in the disk atmosphere or in a dense part of the outflow,  
or possibly from ongoing formation of the molecules.  Both the formation/heating
circumstances and the system geometry remain poorly constrained.


\section*{Acknowledgments}

We thank Ted Bergin and Colette Salyk for very helpful discussions
concerning the origins of hot dense molecular emission 
in protostellar environments,
Russel White for revisiting old observations, and Kunal Mooley
for a thorough literature search.  John Carpenter, Joan Najita, and an
anonymous referee provided thought-provoking comments on the manuscript.
This research made use of three indispensable databases: SIMBAD,
operated at CDS, Strasbourg, France; NASA's Astrophysics
Data System at the Center for Astrophysics, Harvard University; and
NASA's Infrared Science Archive at the Infrared Processing and Analysis 
Center, California Institute of Technology.

\appendix
Near-infrared spectra of IRAS~04369+2539 and IRAS~05451+0037 
were kindly provided by Kevin Covey, after our initial manuscript submission.
They are based on IRTF/SpeX observations taken on 2 September, 2011
and are illustrated in Figure 5.


\newpage
\begin{figure}[tb]
\begin{center}
\includegraphics[angle=270,width=\textwidth]{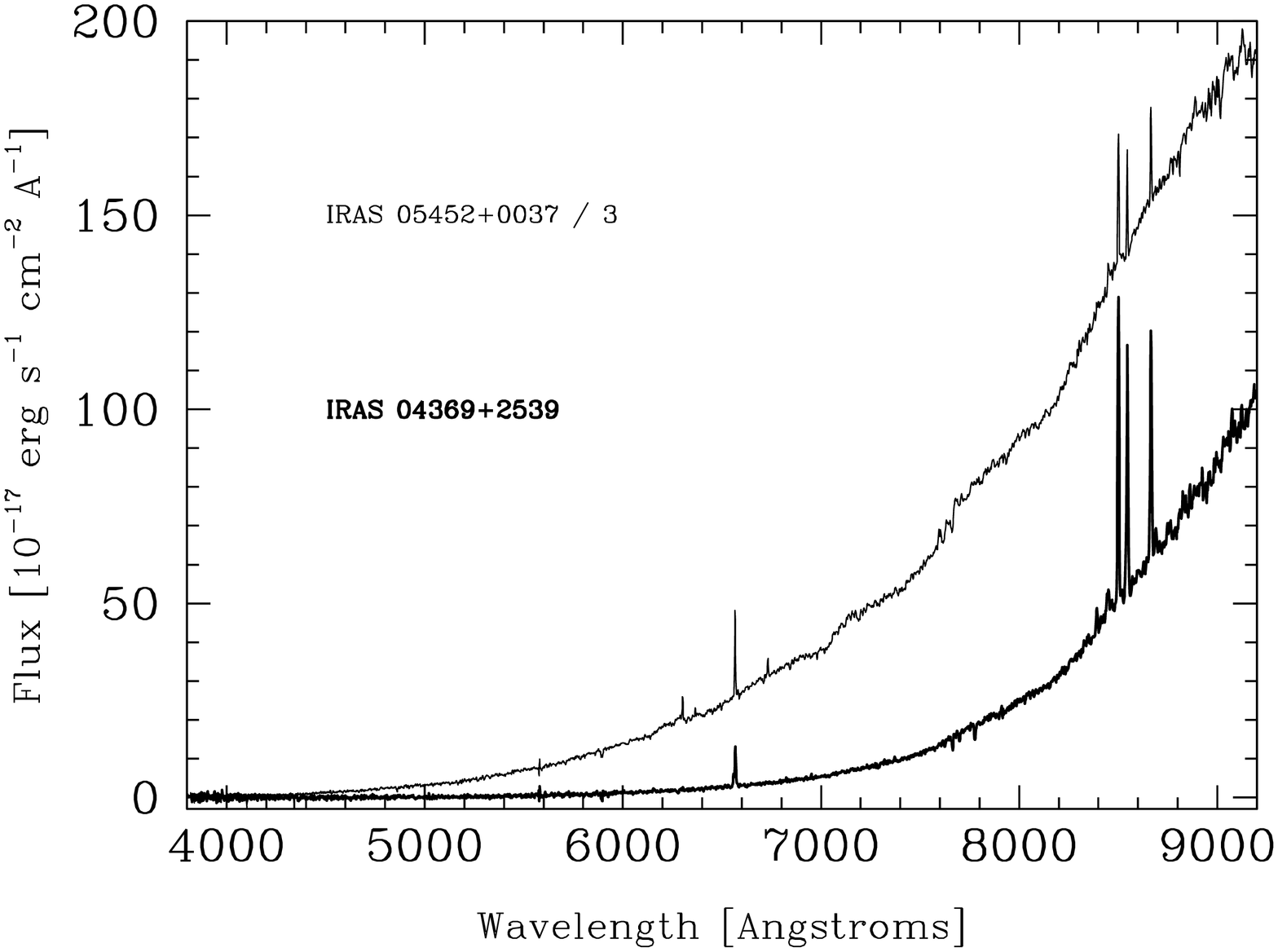}
\end{center} \figcaption{{\it SDSS} spectra of IRAS~04369+2539 
associated with IC2087-IR
(SDSS~J043955.76+254501.9; heavy line, bottom) and 
IRAS~05451+0037 in NGC 2071N 
(SDSS~J054744.99+003841.2; light line, top).
The vertical scale for the IRAS~05451+0037 data is reduced by a factor of 3. 
Both spectra are highly reddened.
\label{fig:fig1}}
\end{figure}

\begin{figure}[tb]
\begin{center}
\plotone{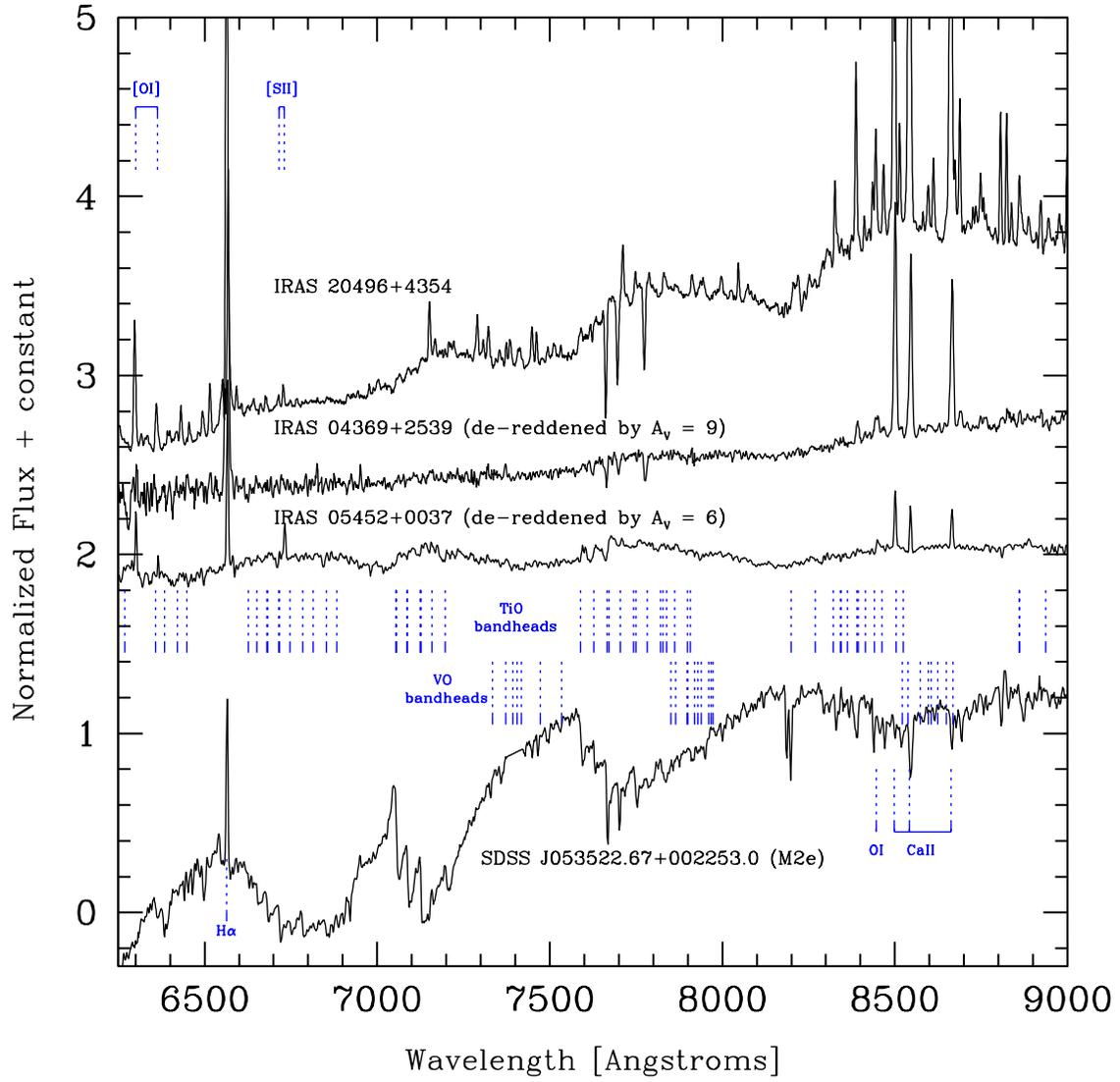}
\end{center} \figcaption{{\it SDSS} spectra of IRAS~04369+2539 
dereddened by $\rm A_V = 9^m$, IRAS~05451+0037 
dereddened by $\rm 6^m$, and 
as comparison objects
the dM2e star SDSS~J053522.67+002253.0 and
the outbursting young star IRAS~20496+4354 (PTF~10nvg; 
Covey et al. 2011), the only other source known to have TiO full band emission. 
The two rows of vertical dotted lines show the wavelengths of the
TiO bandheads tabulated by Valenti et al. (1998) and the
VO bandheads compiled by Kirkpatrick et al. (1991).
When compared to the deep absorption in the dMe star, 
TiO and VO bands are clearly seen fully in emission 
in all three {\it IRAS} sources, strongest in IRAS~20496+4354 
and announced here for the first time in IRAS~04369+2539 and IRAS~05451+0037.
Several other permitted and forbidden lines that appear in emission
in one or more of the IRAS sources are also noted.
\label{fig:fig2}}
\end{figure}

\begin{figure}[tb]
\begin{center}
\includegraphics[angle=270,width=\textwidth]{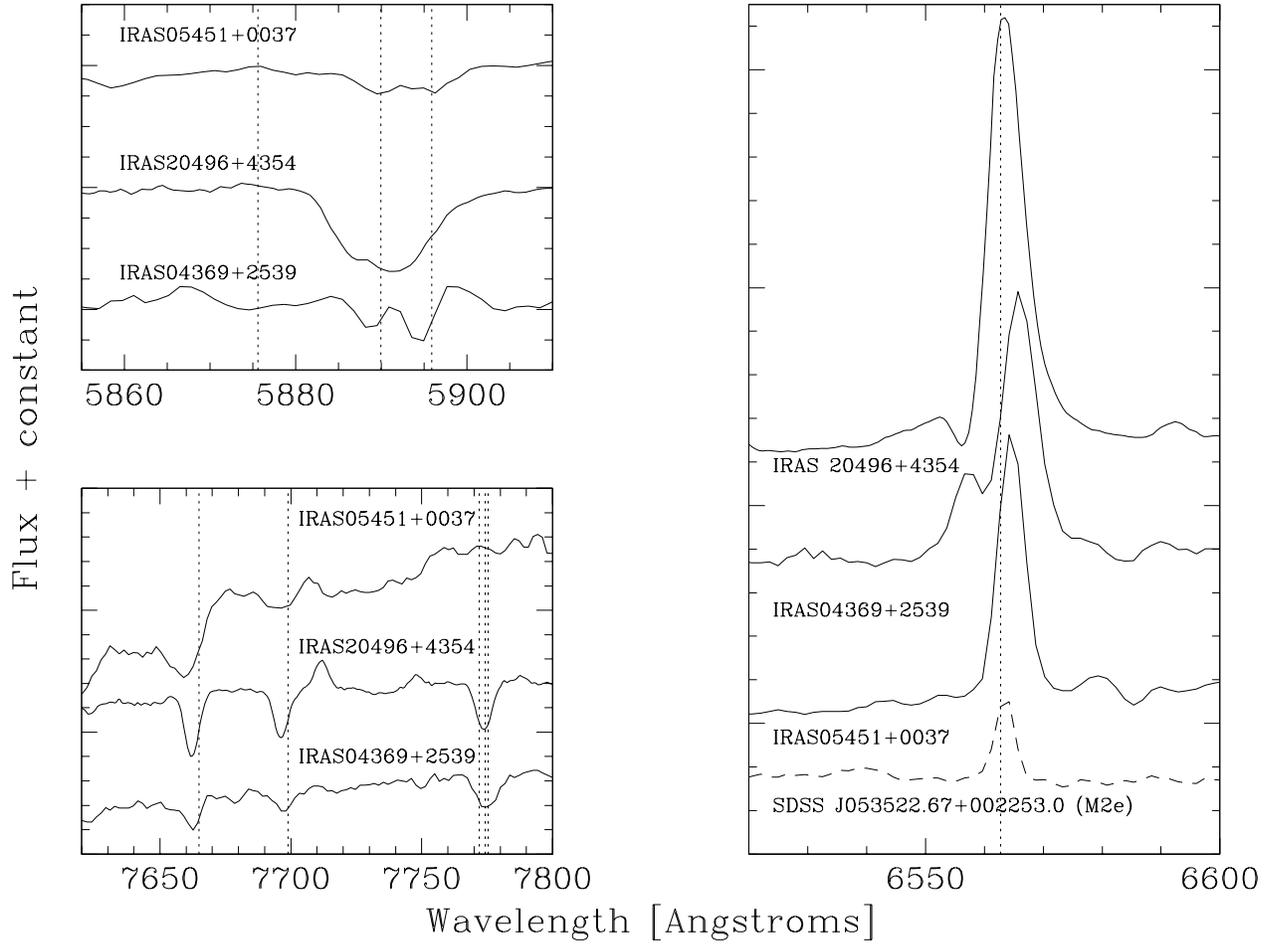}
\end{center} \figcaption{Spectral regions around the 
\ion{Na}{1} D doublet (top left, also including \ion{He}{1}),
and the \ion{K}{1} 7665/7669 doublet and \ion{O}{1} 7773 triplet (bottom left) 
lines which are all seen 
in absorption against the TiO emission pseudo-continuum, as well as 
the H$\alpha$ line (right) seen in pure emission. 
Vacuum wavelengths of the SDSS data are corrected. 
The data for IRAS~04369+2539 and IRAS~05451+0037  
are again compared with those for IRAS~20496+4354 (PTF~10nvg;
Covey et al. 2011). 
The H$\alpha$ emission line for 
the dM2e star SDSS~J053522.67+002253.0 
is also shown in the right panel.
Vertical dotted lines indicate line rest wavelengths with the  
data in the heliocentric standard frame of rest.  The \ion{K}{1} lines
of all three sources and the \ion{Na}{1} D lines in two of the three
appear to have asymmetric blueshifted absorption, consistent with
formation in a wind.   The H$\alpha$ line in the three YSOs
is stronger and broader than typical of dMe
stars and exhibits a range of morphologies, consistent with accretion/outflow
signatures of young stars.
\label{fig:fig3}}
\end{figure}

\begin{figure}[tb]
\begin{center}
\includegraphics[angle=270,totalheight=0.5\textheight]{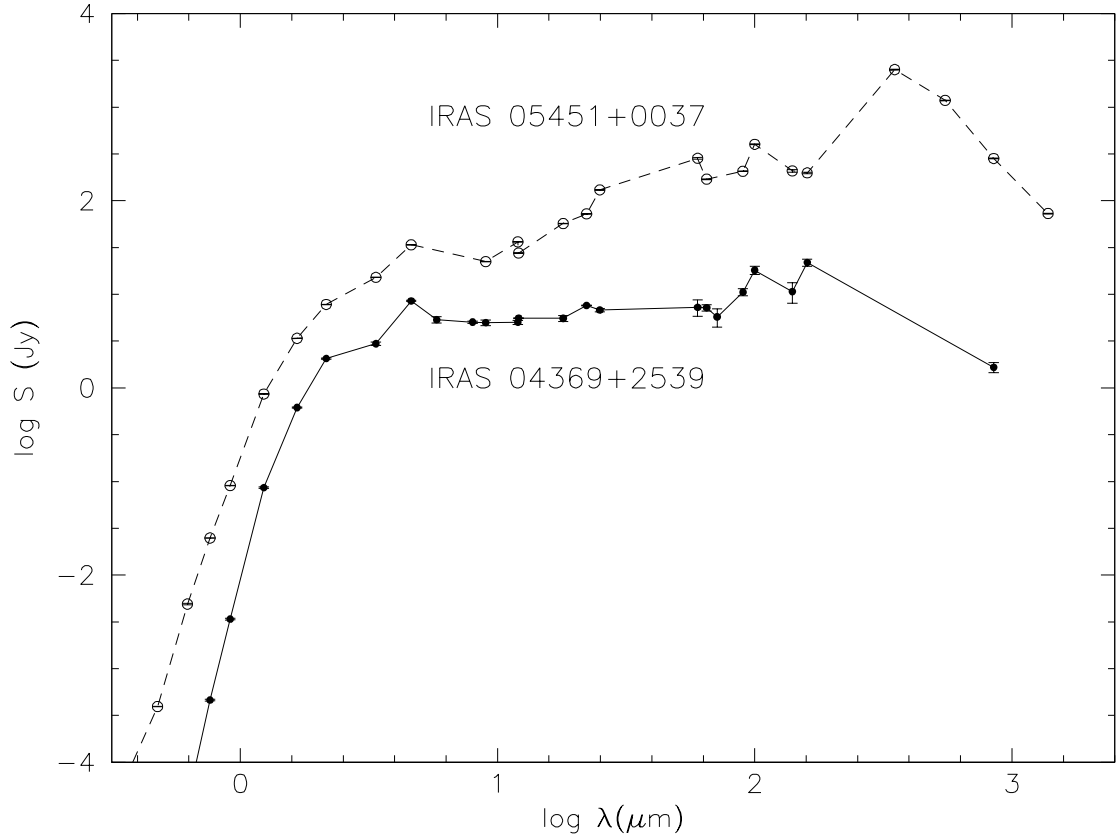}
\end{center} \figcaption{Observed broad-band spectral energy distributions
of IRAS~04369+2539, IC 2087-IR (filled symbols, solid line) and IRAS~05451+0037
(open symbols, dashed line). The plot for IRAS~05451+0037 is raised 
by a value of 1.0 to separate the curves. The data are from
{\it SDSS, 2MASS, WISE, Spitzer, IRAS, Akari, Planck} and JCMT/SCUBA.
Both sources are optically faint and far-infrared bright, and have
spectral energy distributions consistent with those for Class I/II YSOs.
\label{fig:fig4}}
\end{figure}


\begin{figure}[tb]
\begin{center}
\includegraphics[angle=0,totalheight=0.8\textheight]{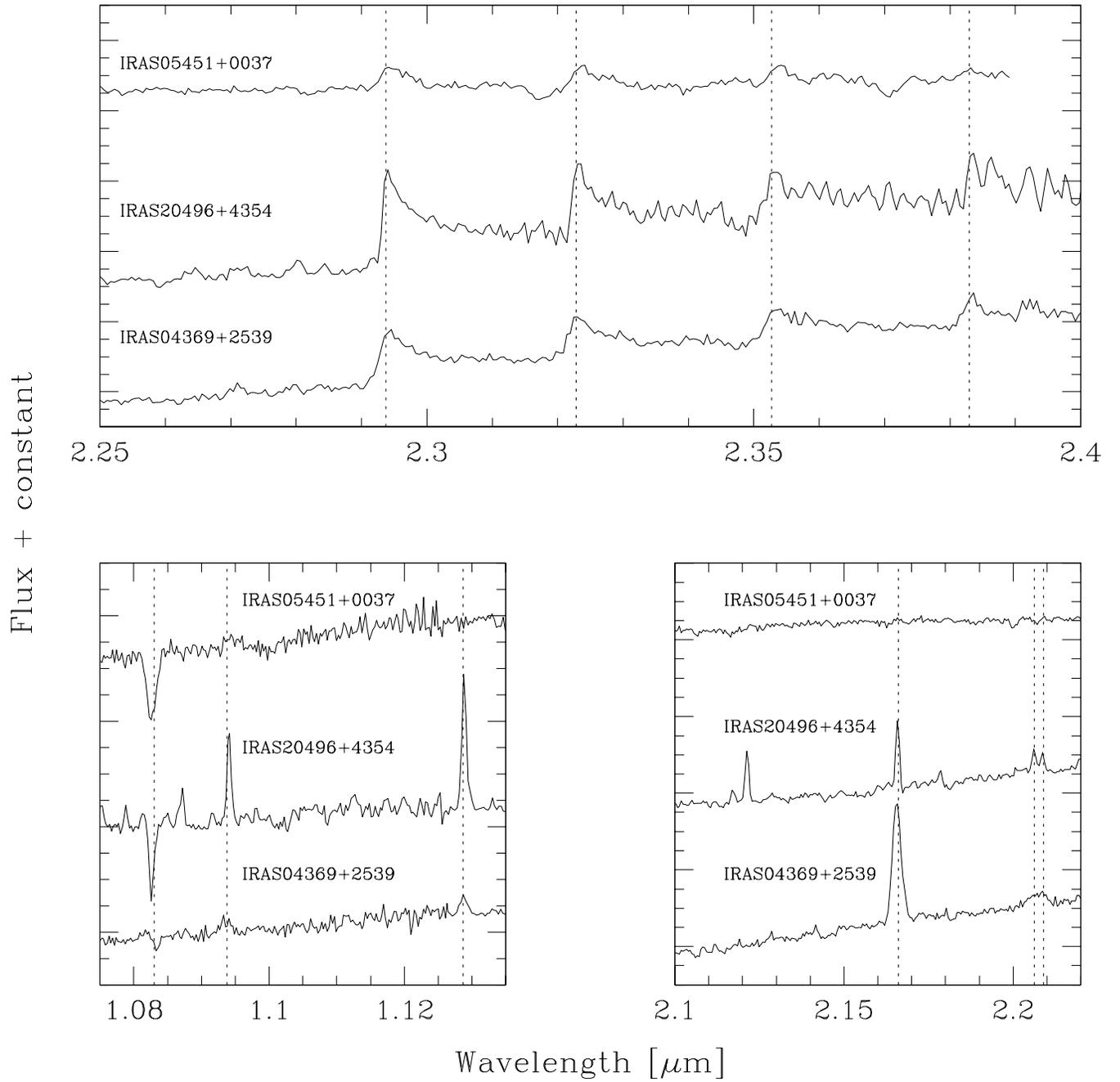}
\end{center} \figcaption{
Infrared spectra of IRAS~04369+2539 and IRAS~05451+0037, kindly provided
by Kevin Covey based on IRTF/SpeX observations.  
Similar to the comparison source
IRAS~20496+4354 (PTF~10nvg; Covey et al. 2011), the newly identified TiO/VO
emitting sources also exhibit prominent CO band emission (top panel).
Spectral regions around the \ion{He}{1} 1.08 $\mu$m,
\ion{H}{1} 1.09 $\mu$m (Pa$\delta$) and \ion{O}{1} 1.13 $\mu$m 
lines are shown in the bottom left while the bottom right shows the 
\ion{H}{1} 2.16 $\mu$m (Br$\gamma$) and \ion{Na}{1} 2.21 $\mu$m region.
IRAS~05451+0037 is fairly featureless other than the strong  
and broad \ion{He}{1} 1.08 $\mu$m absorption; IRAS~04369+2539 has several
emission lines in common with IRAS~20496+4354. 
\label{fig:fig5}}
\end{figure}

\clearpage
\begin{deluxetable}{rcccccccc}
\tabletypesize{\scriptsize}
\tablewidth{0pc}
\tablecaption{Optical Photometry and Spectrophotometry}
\tablehead{
\colhead{MJD-2400000} &
\colhead{Source} &
\colhead{$u$} &
\colhead{$g$} &
\colhead{$r$} &
\colhead{$i$} &
\colhead{$z$} &
\colhead{$Q_{riz} (R_V = 3.1)$} &
\colhead{$Q_{riz} (R_V = 5.1)$} \\
}

\startdata
\multicolumn{7}{c}{IRAS~04369+2539 (SDSS~J043955.76+254501.9)} \\
\hline
33625 & USNO-B& \nodata &  \nodata & 19.82& 16.03& \nodata & \nodata & \nodata \\
48295 & USNO-B& \nodata & \nodata& 19.15& \nodata& \nodata & \nodata & \nodata \\
52639& SDSS Run 3559& $>$22.12& $>$25.40& 19.98& 17.24& 14.94  & -0.36 & 0.23\\
52944& SDSS spectrum& \nodata & $>$23.60& 20.47& 17.82& \nodata  & \nodata &\nodata  \\

%
\\
\hline
\multicolumn{7}{c}{ IRAS~05451+0037 (SDSS~J054744.99+003841.2)} \\
\hline

33978 & USNO-B& \nodata & 19.47& 15.41& 13.77&  \nodata & \nodata & \nodata \\
47525 & USNO-B&  \nodata& 19.48& 16.17&  \nodata&  & \nodata & \nodata \\
51134& SDSS Run 259& 21.40& 19.76& 16.91& 15.12& 13.68  & -0.15  &  0.22   \\
51138& SDSS Run 287& 21.32& 19.68& 16.88& 15.11& 13.62  & -0.24   & 0.14\\
51139& SDSS Run 297& 21.74& 19.69& 16.92& 15.22& 13.66  & -0.41   & 0.00    \\
52312& SDSS Run 2955& 21.08& 19.73& 16.78& 14.93& 13.62  & 0.08    &  0.42   \\
52314& SDSS Run 2968& 21.21& 19.63& 16.84& 15.08& 13.69  & -0.12   &  0.24   \\
52677& SDSS spectrum& \nodata & 19.75& 16.81& 14.97&  \nodata  & \nodata &  \nodata \\ 
53288& SDSS Run 4874& 21.22& 19.65& 16.84& 15.05& 13.64 & -0.11   & 0.25    \\

\enddata
\bigskip
\tablecomments{The USNO-B magnitudes are derived from digitized photographic plates and have errors estimated at 0.2 mag; it should be noted that these
photographic magnitudes are not on the same photometric system as the
SDSS magnitudes in the rest of the table.
The SDSS photometric errors are 0.15 mag in $u$ and 
$\sim$0.02 mag in $g$, $r$, $i$ and $z$;  the resulting error in
 $Q_{riz}$ is $\sim$0.05 mag.
The magnitudes denoted ``spectrum'' are synthesized
from the spectra shown in Figure 1 (see text)}

\end{deluxetable}
\begin{deluxetable}{rrccccc}
\tabletypesize{\scriptsize}
\tablewidth{0pc}
\tablecaption{Representative Disk/Envelope Models}
\tablehead{
\colhead{} &
\colhead{IRAS~04369+2539} &
\colhead{IRAS~05451+0037} \\
}

\startdata

Assumed Distance (pc)& 140& 400 \\
Integrated SED Luminosity ($\rm L_{\odot}$)& 7& 90 \\
Model Total Luminosity ($\rm L_{\odot}$)& 79& 360 \\
Model Foreground Extinction ($A_V$)& 9& 6\\
$\rm M_{disk} ~~(M_{\odot})$& $2.6 \times 10^{-3}$&  $5.8 \times 10^{-3}$ \\
$\rm M_{env} ~~(M_{\odot})$& $6.3 \times 10^{-8}$& $1.5 \times 10^{-5}$\\
$\rm dM_{disk}/dt ~~(M_{\odot}/yr)$& $2.3 \times 10^{-7}$&  $2.2 \times 10^{-7}$ \\
$\rm dM_{env}/dt ~~(M_{\odot}/yr)$& -- &  --  \\
\enddata

\end{deluxetable}

\end{document}